\documentstyle[epsfig]{article}
\begin{document}

\title{Tunneling Times and Superluminality:\\
a Tutorial}
\author{Raymond Y. Chiao\thanks{On appointment as a Miller Research 
Professor in the Miller Institute for Basic Research in Science.}\\
Dept. of Physics, Univ. of California\\ Berkeley, CA 
94720-7300, U. S. A.}
\maketitle

\begin{abstract}
Experiments have shown that individual photons penetrate an optical 
tunnel barrier with an effective group velocity considerably greater 
than the vacuum speed of light.  The experiments were conducted with a 
two-photon parametric down-conversion light source, which produced 
correlated, but random, emissions of photon pairs.  The two photons of 
a given pair were emitted in slightly different directions so that one 
photon passed through the tunnel barrier, while the other photon 
passed through the vacuum.  The time delay for the tunneling photon 
relative to its twin was measured by adjusting the path length 
difference between the two photons in a Hong-Ou-Mandel interferometer, 
in order to achieve coincidence detection.  We found that the photon 
transit time through the barrier was smaller than the twin photon's 
transit time through an equal distance in vacuum, indicating that 
the process of tunneling in quantum mechanics is superluminal.  
Various conflicting theories of tunneling 
times are compared with experiment.
\end{abstract} 

\section*{An introduction to an introduction}

Tunneling, the quantum mechanical process by which a particle can 
penetrate a classically forbidden region of space, is one of the most 
mysterious phenomena of quantum mechanics.  Yet it is one of the most 
basic and important processes in Nature, without which we could not 
even exist, for tunneling is involved in the very first step of the 
nuclear reaction, $p + p \rightarrow d + e + \overline{\nu_{e}}$, 
which powers the Sun, the source of energy for life on the Earth.

It may seem that all that we want to know about tunneling is now well 
understood, given the present mature development of quantum theory.  
The probability for a particle to penetrate a tunnel barrier is 
calculated in elementary textbooks on quantum mechanics.  Not only has 
tunneling now been observed in many settings, but it is also the 
physical basis for many useful devices.  However, there remains an 
open problem concerning the duration of the tunneling process, i.e., 
the question ``How long does is take for the particle to tunnel 
across the barrier?''  This question is still the subject of much 
controversy, since numerous theories contradict each other in their 
predictions for ``the'' tunneling time.  Indeed, some of these 
theories, most notably Wigner's, predict that this time should be 
superluminal, but others predict that it should be subluminal.  Apart 
from its fundamental interest, a correct solution of this problem is 
important for determining the speed of devices which are based on 
tunneling.

Therefore we decided that it would be useful to perform experiments 
which used the photon as the tunneling particle to {\em measure} this 
time.  In this endeavor, we learned the important lesson that a clear 
{\em operational} definition of the experimental method by which the 
tunneling time is measured is necessary before the above question can 
even be well formulated.  In fact, different operational procedures 
will lead to conflicting experimental outcomes, so that the time or 
duration of a process in quantum physics, such as tunneling, is no 
longer unique, in contrast to the situation in classical physics.

\section*{A brief history of tunneling}

Shortly after the advent of the Schr{\"o}dinger equation, 
Hund~\cite{Hund=1927} first noticed the possibility of the phenomenon 
of tunneling, then known as ``barrier penetration.''  He first came 
across this uniquely quantum phenomenon in a calculation of the 
splitting of the ground state in a double-well potential, such as in 
the ammonia molecule in an ammonia maser (to give a modern example).  
Closely following upon the heels of Hund, 
Nordheim~\cite{Nordheim=1927} also applied the then-recently 
discovered Schr{\"o}dinger equation to the calculation of the 
reflection coefficient of an electron from various kinds of 
interfaces.  For the case of a rectangular potential barrier, he noted 
the remarkable fact that an electron whose energy was insufficient to 
go over the barrier classically, can still tunnel through the barrier 
quantum mechanically, whereas classically it would of course always be 
completely reflected from the barrier.  Thus Nordheim extended the 
case of tunneling between bound states first noticed by Hund to the 
case of tunneling between continuum states.  It is with this latter 
case that we shall be concerned with here when we examine the problem 
of tunneling times.

In a rapid sequence of developments in 1928, quantum tunneling was 
applied both to nuclear physics and to solid state physics, i.e., to 
an explanation of $\alpha$ radioactive decay of nuclei, and to an 
explanation of field emission of electrons from a metallic surface.  
As a prelude to these important developments, 
Oppenheimer~\cite{Oppenheimer=1928} performed a correct (but 
unilluminating) calculation of the rate of ionization of a hydrogen 
atom by an external field.  Following this, Gamow~\cite{Gamow=1928}, 
and independently, Gurney and Condon~\cite{Condon=1928} applied the 
newly understood phenomenon of tunneling to explain the enormous range 
of $\alpha$-decay rates of radioactive nuclei.  In an important 
application to solid state physics, Fowler and 
Nordheim~\cite{Fowler=1928} performed a tunneling calculation for the 
rate of emission of electrons into the vacuum due to an intense 
electric field applied to the surface of a metal, then commonly known 
as ``field emission.''  In 1934 Zener~\cite{Zener=1934} applied the 
idea of tunneling to the breakdown of insulators, which was in turn 
applied to Zener diodes.

After the second World War, various striking new tunneling phenomena 
were discovered, and new devices based on tunneling were invented.  In 
1957 Esaki invented the tunnel diode, in which the tunneling of 
electrons and holes based on internal field emission across a narrow 
depletion layer in a heavily doped germanium $p$-$n$ junction led 
to an anomalous I-V characteristic for the junction: There appeared a 
negative resistance region in its I-V curve.  Due to the separation of 
opposite-signed charges on opposite sides of the junction, a large 
internal electric field appeared across the junction.  This resulted 
in a bending of the conduction and valence bands of the semiconductor 
in the depletion region.  The tunneling of electrons and holes across 
this region accounted for the initial increase of current as the bias 
voltage across the junction was initially increased from zero.  
However, as this voltage was increased to a value comparable to the 
band gap of germanium, the tunneling current decreased.  This was due 
to the decrease in the density of permissible final states for the 
electrons and holes when they ended up inside the band gap.  There 
resulted a decrease in transmitted current across the junction on 
account of the Bragg reflection of the electron and holes from the 
periodic germanium lattice, when their momenta approached that of the 
lattice.  Hence an anomalous negative resistance region appeared in 
the I-V curve, which allowed the construction of high-frequency 
electronic oscillators based on the tunnel diode.

In 1962 Josephson predicted the existence of a tunneling 
supercurrent which traversed a gap separating two superconductors.  
This superconducting tunnel effect was confirmed experimentally by 
Giaever and others in Josephson junctions consisting of 
superconducting thin films separated by a thin oxide barrier.  In 
1982 Binnig and Rohrer applied electron tunneling to the invention 
of the scanning tunneling microscope.  By monitoring the tunneling 
current between a sharp metallic needle (etched so that a single 
atom resides at its tip) and a sample surface as the position of the 
needle was scanned over the surface, an image of the surface, where 
individual atoms were resolved, was made possible.  Earlier the 
field-emission microscope invented by Mueller had also allowed the 
imaging of individual atoms near the tip of an etched metallic 
needle by imaging the tunneling electrons which had escaped due to 
field emission from the tip onto a distant screen, but this earlier 
microscope had only a limited usefulness.

\section*{Tunneling time experiments at Berkeley}

The many manifestations of tunneling and the many applications to 
devices strongly motivated us to examine the unsolved tunneling time 
problem experimentally.  It is very important in attacking this 
problem to state precisely at the outset the {\it operational} 
definition of the quantity being measured.  For the tunneling time for 
our experiments, this definition is based on the following {\it 
Gedankenexperiment} (see Fig.~1).
\begin{figure}
\centerline{\psfig{figure=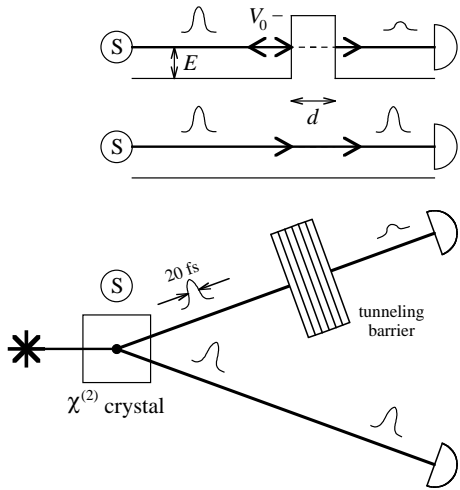,width=6cm}} 
\caption{(Top): {\it Gedankenexperiment} to measure the tunneling time by 
means of two particles simultaneously emitted from a source S and detected 
by two equidistant detectors. (Bottom): Realization by a spontaneous 
parametric down-conversion source S, in which a parent photon decays into 
two daughter photons.} 
\end{figure}  
Suppose that a single parent particle (a photon) decays into two 
daughter particles (two photons), as in a radioactive decay.  Suppose 
further that these two daughter particles have the same speed in the 
vacuum (i.e., $c$ in our case), and that they were detected by means 
of two detectors placed at equal distances from the point of decay.  
There result simultaneous clicks at the two detectors, e.g., two 
Geiger counters, which could then be registered in a coincidence 
counter.  Now suppose that we place a tunnel barrier in the path of 
one of the daughter photons.  (The other daughter photon continues to 
travel unimpeded through the vacuum.)  Of course, this would greatly 
diminish the coincidence count rate.  However, whenever a tunneling 
event does occur, the {\it difference} in the time of arrival of the 
two daughters, as measured by the difference in the time of the clicks 
of the two detectors, constitutes a precise definition for the 
tunneling time.

The question concerning the superluminality of the tunneling process 
can now also be precisely stated.  Does the click of the detector 
which registers the arrival of the photon which traversed the tunnel 
barrier go off {\it earlier} or {\it later} (on the average) than the 
click of the detector which registers the arrival the photon which 
traversed the vacuum?  If the tunnel barrier had simply been a thin 
piece of transparent glass, then the answer would obviously be 
``later,'' since the group velocity for a photon inside the glass 
would be {\it less} than the speed of light, and the group delay for 
the photon traversing the glass relative to that of the vacuum would 
be {\it positive}.  However, if, as some tunneling-time theories 
predict, the tunneling process is superluminal, then the 
counterintuitive answer would be ``earlier,'' since the effective 
group velocity for a photon inside the tunnel barrier would be {\it 
greater} than the speed of light, and the group delay for the photon 
traversing the barrier relative to that of the vacuum would be {\it 
negative}.  Hence it is the {\it sign} of the relative time between 
the clicks in the two detectors which determines whether tunneling is 
subluminal or superluminal.

The reader may ask why relativistic causality is not violated by the 
superluminality of the tunneling process, if it should indeed be 
superluminal.  It has been shown~\cite{Chiao=1997} that special 
relativity does not forbid the $group$ velocity to be faster than 
$c$; only Sommerfeld's $front$ velocity must not exceed $c$.  Also 
remember that due to the uncertainty principle the time of emission 
of the signal photon is not under the experimenter's control.

Presently, the best detectors for photons have picosecond-scale 
response times, which are still not fast enough to detect the 
femtosecond-scale time differences expected in our tunneling-time 
experiment.  Hence it was necessary to utilize a Hong-Ou-Mandel 
interferometer, which has a femtosecond-scale temporal resolution for 
measuring the time difference between the travel times of the two 
photons traversing the two arms of the interferometer.  By placing the 
tunnel barrier in one of these arms, a precise measurement of the 
delay due to tunneling could then be performed.
  
The tunnel barrier used in our experiments was a dielectric mirror in 
which periodic layers of alternately high and low index media produce 
a photonic band gap at the first Brillouin zone edge.  The problem of 
photon propagation in this periodic structure is analogous to that of 
the Kronig-Penney model for electrons propagating inside a crystal.  
In particular, near the midgap point on the first Brillouin zone edge, 
there exists due to Bragg reflection inside the periodic structure an 
evanescent (i.e., exponential) decay of the transmitted wave 
amplitude, which is equivalent to tunneling.  Note that this Bragg 
reflection effect is completely analogous to the one occurring in the 
Esaki tunnel diode mentioned above.  One important feature of this 
kind of tunnel barrier is the fact that it is nondispersive near 
midgap, and therefore there is little distortion of the tunneling wave 
packet.

\section*{Tunneling Time Theories}

Another strong motivation for performing experiments to measure the 
tunneling time was the fact that there were many conflicting theories 
for this time (see the reviews by Hauge and 
St{\o}vneng~\cite{Hauge=1989}, by Landauer and 
Martin~\cite{Landauer=1994}, and by Chiao and 
Steinberg~\cite{Chiao=1997}).  It suffices here to list the three 
main contenders:

(1) The Wigner time (i.e., ``phase time'' or ``group delay'').

(2) The B\"{u}ttiker-Landauer time (i.e., ``semiclassical time'').

(3) The Larmor time (with B\"{u}ttiker's modification).

The Wigner time calculates how long it takes for the peak of a wave 
packet to emerge from the exit face of the tunnel barrier relative to 
the time the peak of the incident wave packet arrives at the entrance 
face.  Since the peak of the wave packet in the Born interpretation is 
the point of highest probability for a click to occur (see the above 
{\it Gedankenexperiment}), it is natural to expect this to be the 
relevant time for our experiments.  This calculation is based on an 
asymptotic treatment of tunneling as a scattering problem, and 
utilizes the method of stationary phase to calculate the position of 
the peak of a wave packet.  The result is simple: this tunneling time 
is the derivative of the phase of the tunneling amplitude with respect 
to the energy of the particle.

The B\"{u}ttiker-Landauer time is based on a different {\it 
Gedankenexperiment}.  Suppose that the height of the tunnel barrier is 
perturbed sinusoidally in time.  If the frequency of the perturbation 
is very low, the tunneling particle will see the instantaneous height 
of the barrier, and the transmission probability will adiabatically 
follow the perturbation.  However, as one increases the frequency of 
the perturbation, at some characteristic frequency the tunneling 
probability will no longer be able to adiabatically follow the rapidly 
varying perturbation.  It is natural to define the tunneling time as 
the inverse of this characteristic frequency.  The result is again 
simple: for opaque barriers, this tunneling time is the distance 
traversed by the particle (i.e., the barrier width $d$) divided by the 
absolute value of the velocity of the particle $|v|$.  (In the 
classically forbidden region of the barrier, this velocity is 
imaginary, but its characteristic size is given by the absolute 
value).

The Larmor time is based on yet another {\it Gedankenexperiment}.  
Suppose that the tunneling particle had a spin magnetic moment (e.g., 
the electron).  Suppose further that a magnetic field were applied to 
region of the barrier, but only to that region.  Then the angle of 
precession of the spin of the tunneling particle is a natural measure 
of the tunneling time.  However, B\"{u}ttiker noticed that in addition 
to this Larmor precession effect, there is a considerable tendency for 
the spin to align itself either along or against the direction of the 
magnetic field during tunneling, since the energy for these two spin 
orientations is different.  The {\it total} angular change of the 
tunneling particle's spin divided by the Larmor precession frequency 
is B\"{u}ttiker's Larmor time.

One consequence of the Wigner time is the Hartman effect: The 
tunneling time saturates for opaque barriers, and approaches for 
large $d$ a limiting value given by the uncertainty principle, 
$\hbar/(V_{0}-E)$.  The apparent superluminality of tunneling is a 
consequence of this effect, since as $d$ is increased, there is a 
point beyond which the saturated value of the tunneling time is 
exceeded by the vacuum traversal time $d/c$, and the particle 
appears to have tunneled faster than light.

By contrast, the B\"{u}ttiker-Landauer theory predicts a tunneling 
time which increases linearly with $d$ for opaque barriers, as one 
would expect classically.  For a rectangular barrier with a height 
$V_{0}<<mc^{2}$, the effective velocity $|v|$ is always less than $c$.  
However, for the periodic structure which we used in our experiment, 
the effective velocity $|v|$ at midgap is infinite, which is a 
behavior even more superluminal than that predicted by the Wigner 
time.  This fact makes it easy to distinguish experimentally between 
these two theories of the tunneling time.  However, we hasten to add 
that the B\"{u}ttiker-Landauer time may not apply to our experimental 
situation, as the {\it Gedankenexperiment} on which it is based is 
quite different from the one relevant to our experiment.

B\"{u}ttiker's Larmor time predicts a tunneling time which is 
independent of $d$ for thin barriers, but which asymptotically 
approaches a linear dependence on $d$ in the opaque barrier limit, 
where it coincides with the B\"{u}ttiker-Landauer time.  In our first 
experiment it was impossible to distinguish experimentally between 
this time and the Wigner time.  Only in our second experiment could 
these two theories be clearly distinguished from one another.

\section*{Details of the Berkeley Experiments}

Spontaneous parametric down-conversion was the light source used in 
our experiments~\cite{Steinberg=1993,Steinberg=1995}.  An 
ultraviolet (UV) beam from an argon laser operating at a wavelength of 
351 nm was incident on a crystal of potassium dihydrogen phosphate 
(KDP), which has a ${\chi}^{(2)}$ nonlinearity.  During the process of 
parametric down-conversion inside the crystal, a rainbow of many 
colors was generated in conical emissions around the ultraviolet laser 
beam, in which one parent UV photon broke up into two daughter 
photons, conserving energy and momentum.  The KDP crystal was cut with 
an optic axis oriented so that the two degenerate (i.e., equal energy) 
daughter photons at a wavelength of 702 nm emerged at a small angle 
relative to each other.  We used two pinholes to select out these two 
degenerate photons.  The size of these pinholes determined the 
bandwidth of the light which passed through them, and the resulting 
single-photon wavepackets had temporal widths around 20 fs and a 
bandwidth of around 6 nm in wavelength.

The tunnel barrier consisted of a dielectric mirror with eleven 
quarter-wavelength layers of alternately high index material (titanium 
oxide with $n=2.22$) and low index material (fused silica with 
$n=1.45$).  The total thickness of the eleven layers was 1.1 ${\rm 
{\mu m}}$.  This implied an {\it in vacuo} traversal time across the 
structure of 3.6 fs.  Viewed as a photonic bandgap medium, this 
periodic structure had a lower band edge located at a wavelength of 
800 nm and an upper band edge at 600 nm.  The transmission coefficient 
of the two photons which were tuned near midgap (700 nm) was 1\%.  
Since the transmission had a broad minimum at midgap compared to the 
wave packet bandwidth, there was little pulse distortion.  The Wigner 
theory predicted at midgap a tunneling delay time of around 2 fs, or 
an effective tunneling velocity of $1.8c$.  The B\"{u}ttiker-Landauer 
theory predicted at midgap an infinite effective tunneling velocity, 
which implies a zero tunneling time.

To achieve the femtosecond-scale temporal resolutions necessary for 
measuring the tiny time delays associated with tunneling, we brought 
together these two photons by means of two mirrors, so that they 
impinged simultaneously at a beam splitter before they were detected 
in coincidence by two Geiger-mode silicon avalanche photodiodes.  
There resulted a narrow null in the coincidence count rate as a 
function of the relative delay between the two photons, a 
destructive interference effect first observed by Hong, Ou, and 
Mandel~\cite{Hong=1987}.  The narrowness of this coincidence 
minimum, combined with a good signal-to-noise ratio, allowed a 
measurement of the relative delay between the two photons to a 
precision of $\pm 0.2$ fs.

A simple way to understand this two-photon interference is to apply 
Feynman's rules for the interference of indistinguishable processes.  
Consider two photons impinging simultaneously on a 50/50 beam splitter 
followed by two detectors in coincidence detection.  When two 
simultaneous clicks occur at the two detectors, it is impossible even 
in principle to tell whether both photons were reflected by the beam 
splitter or whether both photons were transmitted through the beam 
splitter.  In this case, Feynman's rules tell us to add the 
probability amplitudes for these two indistinguishable process, and 
then take the absolute square to find the probability.  Thus the 
probability of a coincidence count to occur is given by $|r^{2} + 
t^{2}|^{2}$, where $r$ is the complex reflection amplitude for one 
photon to be reflected, and $t$ is the complex transmission amplitude 
for one photon to be transmitted.  For a lossless beam splitter, 
time-reversal symmetry leads to the relation $t={\pm}ir$.  
Substituting this into the expression for the coincidence probability, 
and using the fact that $|r|=|t|$ for a 50/50 beam splitter, we find 
that this probability vanishes.  Thus the two photons must always pair 
off in the same (random) direction towards only one of the two 
detectors, an effect which arises from the bosonic nature of the 
photons. 

A schematic of the apparatus we used to measure the tunneling time is 
given in Fig.~2.  
The delay between the two daughter photons was 
adjustable by means of the ``trombone prism" mounted on a Burleigh 
inchworm system, and was measured by means of a Heidenhein encoder 
with a 0.1 ${\rm {\mu}m}$ resolution.  A {\it positive} sign of the 
delay due to a piece of glass was determined as corresponding to a 
motion of the prism {\it towards} the glass.  The multilayer coating 
of the dielectric mirror (i.e., the tunnel barrier) was evaporated on 
only half of the glass mirror substrate.  This allowed us to translate 
the mirror so that the beam path passed either through the tunnel 
\begin{figure}
\centerline{\psfig{figure=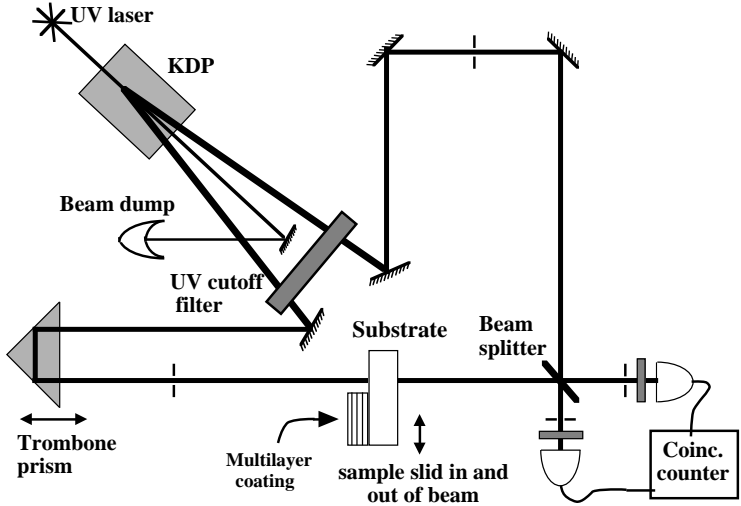,width=10cm}} 
\caption{Schematic of the 
Berkeley experiments to measure the tunneling time.}
\end{figure}
barrier in an actual measurement of the tunneling time, or through the 
uncoated half of the substrate in a control experiment.  In this way, 
one could obtain data with and without the barrier in the beam, i.e., 
a direct comparison between the delay through the tunneling barrier 
and the delay for traversing an equal distance in air.  The normalized 
data obtained in this fashion is shown in Fig.~3(a), with the barrier
\begin{figure}
\centerline{\psfig{figure=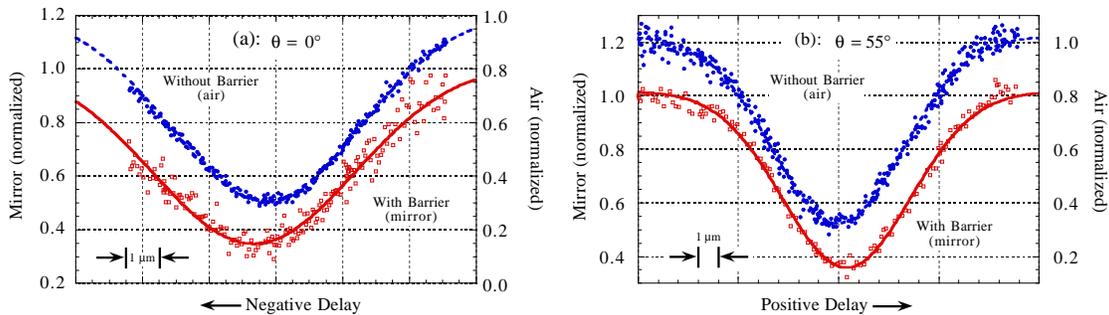,width=15cm}} 
\caption{(a) Coincidence rate vs delay (i.e., the position of the 
trombone prism in Fig.~2) with and without the tunnel barrier 
(mirror) in the beam path at normal incidence.  (b) The same with the 
mirror tilted towards Brewster's angle.} 
\end{figure} 
oriented at normal incidence ($\theta=0^{\circ}$).  Note that the 
coincidence minimum with the tunnel barrier in the beam is shifted to 
a {\it negative} value of delay relative to that without the barrier 
in the beam.  This negative shift indicates that the tunneling delay 
is superluminal.

To double-check the sign of this shift, which is crucial for the 
interpretation of superluminality, we tilted the mirror towards 
Brewster's angle for the substrate ($\theta=56^{\circ}$), where there 
is a very broad minimum in the reflection coefficient as a function of 
angle.  Near Brewster's angle this minimum is so broad that it is not 
very sensitive to the difference between the high and low indices of 
the successive layers of dielectrics.  Thus to a good approximation, 
the reflections from all layers vanish simultaneously near this angle.  
Hence the Bragg reflection responsible for the band gap disappears, 
and the evanescent wave behavior and the tunneling behavior seen near 
normal incidence disappears.  The dielectric mirror should then behave 
like a thin piece of transparent glass with a {\it positive} delay 
time relative to that of the vacuum.  Detailed calculations not using the 
above approximations also show that at $\theta=55^{\circ}$, the sign 
of the shift should indeed revert to its normal positive value.

The data taken in p-polarization at $\theta=55^{\circ}$ is shown in 
Fig.~3(b).  The reversal of the sign of the shift is clearly seen.  
Therefore one is confronted with a choice of the data either in 
Fig.~3(a) or in Fig.~3(b) as showing a superluminal shift.  Since we 
know that the delay in normal dielectrics as represented by Fig.~3(b) 
should be subluminal, this implies that the tunneling delay in Fig.~3(a) 
should be superluminal.  Therefore the data in Fig.~3(a) implies 
that after traversing the tunnel barrier, the peak of a photon wave 
packet arrived $1.47 \pm 0.21$ fs {\it earlier} than it would had it 
traversed only vacuum.

Another reason for tilting the mirror is that one can thereby 
distinguish between the Wigner time and B\"{u}ttiker's Larmor time, as 
they differ considerably in the region near the band edge, which 
occurs near Brewster's angle.  This can be seen in Fig.~4, where
\begin{figure}
\centerline{\psfig{figure=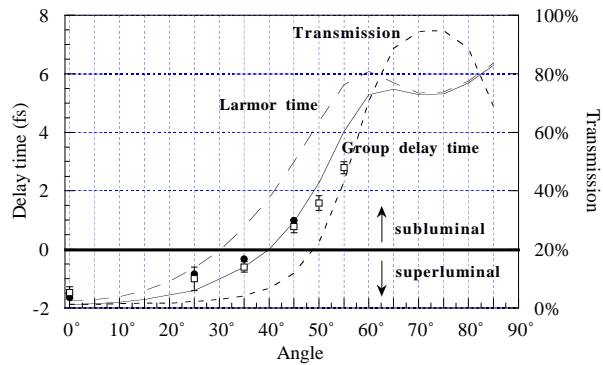,width=8cm}} 
\caption{Temporal shifts of the minima seen in data such as in Fig.~3 
plotted as a function of incidence angle, compared with the 
theoretical predictions of the Wigner time (solid curve) and of 
B\"{u}ttiker's Larmor time (long-dashed curve).  The transmission vs 
incidence angle (short-dashed line).  All curves are for 
p-polarization.  Solid circles and open squares denote data from two 
different tunnel barrier samples.}
\end{figure} 
there is a considerable divergence as the band edge is approached 
between the solid line representing the theoretical prediction of the 
Wigner time, and the long-dashed line representing that of 
B\"{u}ttiker's Larmor time.  The data points in Fig.~4 seem to rule 
out B\"{u}ttiker's Larmor time (although again we hasten to add that 
this theory may not apply to our experiment).  The agreement with 
Wigner's theory is better, but there are discrepancies which are not 
understood.

Other experiments confirming the superluminality of tunneling have 
been performed in Cologne, Florence, and 
Vienna~\cite{Enders=1993,Ranfagni=1993,Spielmann=1994}.  The Cologne 
and Florence groups performed microwave experiments, and the Vienna 
group performed a femtosecond laser experiment.  All these groups have 
confirmed the Hartman effect.  One of these 
groups~\cite{Heitmann=1994} has claimed to have sent Mozart's 40th 
symphony at a speed of $4.7c$ through a microwave tunnel barrier 114 
mm long consisting of a periodic dielectric structure similar to our 
dielectric mirror.  However, the further implication that their 
experiment represents a violation of causality is in our opinion 
unfounded \cite{Chiao=1997}.  

Recently, an experiment indicating the simultaneous existence of two 
different tunneling times was performed in 
Rennes~\cite{Balcou=1997}.  In frustrated total reflection (FTIR), 
the tunneling of photons through an air gap occurs between two glass 
prisms when a light beam is incident upon this gap beyond the 
critical angle.  The Rennes group observed in FTIR both a lateral 
{\it displacement} of the tunneling beam of light and an angular 
{\it deflection} of this beam.  These two effects could be 
interpreted as evidence for two different tunneling times that 
simultaneously occurred in the same tunneling barrier.  The lateral 
displacement is related to the Wigner time, and the angular 
deflection is related to the B\"{u}ttiker-Landauer time.  As 
evidence for this, they cited the saturation of the beam 
displacement (the Hartman effect), and the linear increase of the 
beam deflection, as the gap was increased.

\section*{Conclusions}

The experiments at Berkeley and elsewhere thus indicate that the 
tunneling process is superluminal.  In our opinion, this does not 
imply that one can communicate faster than $c$, despite claims to 
the contrary by Heitmann and Nimtz~\cite{Heitmann=1994}.  The group 
velocity cannot be identified as the signal velocity of special 
relativity, by which a cause is connected to its effect.  Rather, it 
is Sommerfeld's front velocity which exclusively plays this role.  
However, even if one were to define the group velocity as a 
``signal'' velocity, no causal loop paradoxes can arise 
\cite{Garrison=1998}.

Although the controversies amongst the various tunneling theories 
have not yet been fully resolved by experiment, a good beginning has 
been made in this direction.  In particular, it is now clear that 
one cannot rule out the Wigner time simply on the grounds that it 
yields a superluminal tunneling time.  It also appears that there 
may exist more than one tunneling time.  Hopefully, the mysterious role 
of time in quantum mechanics will be elucidated by these studies.

\section*{Acknowledgments}

I would like to thank Prof.~Rodolfo Bonifacio for inviting me to 
give the opening lecture of this conference on the mysteries, 
puzzles, and paradoxes of quantum physics.  I would also like to 
thank Dr.~Paul Kwiat and Prof.~Aephraim Steinberg for their 
collaboration on the Berkeley experiments, and for many helpful 
discussions.  This work was supported by the ONR under Grant 
No.~N000149610034.


\begin{thebibliography}{99}

\bibitem{Hund=1927}
F. Hund, Zeitschrift f{\"u}r Physik {\bf 43} (1927) 805.
 
\bibitem{Nordheim=1927}
L. Nordheim, Zeitschrift f{\"u}r Physik {\bf 46} (1927) 833.
 
\bibitem{Oppenheimer=1928}
J. R. Oppenheimer, Phys. Rev. {\bf 31} (1928) 66; Proc. Nat. Acad. 
Sci. {\bf 14} (1928) 363.
 
\bibitem{Gamow=1928}
G. Gamow, Zeitschrift f{\"u}r Physik {\bf 51} (1928) 204; {\bf 52} 
(1928) 510.
 
\bibitem{Condon=1928}
R. W. Gurney and E. U. Condon, Nature {\bf 122} (1928) 439; Phys. 
Rev. {\bf 33} (1929) 127.
 
\bibitem{Fowler=1928}
R. H. Fowler and L. Nordheim, Proc.  Roy.  Soc.  (London) {\bf A119} 
(1928) 173.

\bibitem{Zener=1934}
C. Zener, Proc. Roy. Soc. (London) {\bf 145} (1934) 523.

\bibitem{Chiao=1997}
R. Y. Chiao and A. M. Steinberg, in {\it Progress in Optics} XXXVII, E. 
Wolf, ed., (Elsevier, Amsterdam, 1997), p.  345.

\bibitem{Hauge=1989}
E. H. Hauge and J. A. St{\o}vneng, Rev.  Mod.  Phys.  {\bf 
61} (1989) 917.

\bibitem{Landauer=1994}
R. Landauer and Th. Martin, Rev. Mod. Phys. {\bf 66} (1994) 217.

\bibitem{Steinberg=1993}
A. M. Steinberg, P. G. Kwiat, and  R. Y. Chiao, Phys.  Rev.  
Lett. {\bf 71}(1993) 708.

\bibitem{Steinberg=1995}
A. M. Steinberg, and R. Y. Chiao, Phys. Rev. {\bf 51}(1995) 
3525.

\bibitem{Hong=1987}
C. K. Hong, Z. Y. Ou and L. Mandel, Phys.  Rev.  Lett.  {\bf 
59} (1987) 2044.

\bibitem{Enders=1993}
A. Enders and G. Nimtz, J. Phys.  I France {\bf 3} (1993) 1089.

\bibitem{Ranfagni=1993}
A. Ranfagni, P. Fabeni, G.P. Pazzi and D. Mugnai, Phys.  Rev.  
E {\bf 48}(1993) 1453.

\bibitem{Spielmann=1994} 
Ch. Spielmann, R. Szip\"{o}cs, A. Stingl 
and F. Krausz, Phys.  Rev.  Lett.  {\bf 73}(1994) 2308.

\bibitem{Heitmann=1994}
W. Heitmann and G. Nimtz, Phys. Lett. A {\bf 196}(1994) 154.

\bibitem{Balcou=1997}
Ph. Balcou and L. Dutriaux, Phys. Rev. Lett. {\bf 78}(1997) 851.

\bibitem{Garrison=1998}
J. C. Garrison, M. W. Mitchell, R. Y. Chiao, and E. L. Bolda, Phys. 
Lett. A {\bf 245} (1998) 19.

\end{thebibliography}
\end{document}